\documentclass[12pt]{article} 
\usepackage{amsfonts,amsmath,epsfig} 

\begin{document}
\title{\noindent \textbf{Explicit secular equations for piezoacoustic 
surface waves: \\Shear-Horizontal modes.\\ 
}}

\author{Bernard Collet, Michel Destrade}  
\date{2004} 
\maketitle 

\bigskip

\begin{abstract}
Attention is given to surface waves 
of shear-horizontal modes in piezoelectric crystals permitting the 
decoupling between an elastic in-plane Rayleigh wave and a 
piezoacoustic anti-plane Bleustein-Gulyaev wave.
Specifically, the crystals possess $\bar{4}$ 
symmetry (inclusive of $\bar{4}2$m, $\bar{4}3$m, and 23 classes) 
and the boundary is any plane containing 
the normal to a symmetry plane (rotated $Y$-cuts about the $Z$ axis).
The secular equation is obtained explicitly as a polynomial
not only for the metallized boundary condition but, 
in contrast to previous studies on the subject, also for other types  
of boundary conditions.
For the metallized surface problem, the secular equation is 
a quadratic in the squared wave speed; 
for the un-metallized surface problem, it is a 
sextic in the squared wave speed; 
for the thin conducting boundary problem, it is of degree 
16 in the speed. 
The relevant root of the secular equation can be identified and the 
complete solution is then found (attenuation factors, field profiles, 
etc.).
The influences of the cut angle and of the conductance of the 
adjoining medium are illustrated numerically for GaAs 
($\bar{4}3$m), BaLaGa$_3$O$_7$ ($\bar{4}2$m) and 
Bi$_{12}$GeO$_{20}$ (23).
Indications are given on how to apply the method to crystals with 
222 symmetry.
\end{abstract}


\newpage

\section{Introduction} 

The nature and properties of a piezoacoustic surface wave depend 
heavily on the crystallographic and anisotropic properties of the 
piezoelectric substrate, on the direction of propagation, and on the 
orientation of the cut (boundary) plane. 
For certain choices, the in-plane components of the mechanical 
displacement decouple from the anti-plane component, leading to two 
different types of surface waves namely, the Rayleigh wave, 
elliptically polarized in the plane containing the direction of 
propagation and the normal to the substrate surface, and the 
shear-horizontal (SH) wave, linearly polarized in the direction 
normal to the direction of propagation and parallel to the free 
surface. 
Moreover, either wave or both waves may be coupled to the 
electromagnetic fields. 
Of special interest are the configurations allowing for a 
piezoacoustic SH wave, decoupled from a purely elastic Rayleigh wave. 
Indeed, the former type of wave, also known as 
Bleustein-Gulyaev\cite{Bleu68,Guly69} wave, 
penetrates more deeply into the substrate than the latter type; 
consequently, the acoustic energy is less localized and the power can 
be increased significantly before damage occurs (Tseng\cite{Tsen70})
although, as pointed out by a Referee, there are exceptions to this 
behavior\cite{NaOs97}.
Moreover, a SH wave-based resonator is smaller than a resonator based 
on the propagation of a sagittally polarized surface wave 
(see Kadota and collaborators \cite{KYFN01,KAHI01});
this feature results in a downsizing of design, 
an attractive point for the miniaturization of mobile phones for 
instance, where SAW devices are used as filters.
Koerber and Vogel\cite{KoVo72,KoVo73} identified all the 
cuts and rotations of axes leading to piezoacoustic SH modes; 
they exist for some suitable cuts and transformations in the following 
crystal classes: 2, 23, $\bar{4}3$m, 222, 2mm, 4, $\bar{4}$, 
6, 4mm, 6mm, 32, $\bar{4}2$m, $\bar{6}$m2, 422, and 622.
The main purpose of this paper is to derive explicitly the secular 
equation for piezoacoustic SH surface waves uncoupled from purely 
elastic Rayleigh  waves polarized in a plane of symmetry,
for a crystal in the $\bar{4}$ symmetry class 
(and thus for the $\bar{4}2$m, $\bar{4}3$m, and 23 classes).

Several workers addressed this topic in the wake of the seminal 
papers by Bleustein and by Gulyaev, but explicit results remained 
limited either to propagation in special directions for which one of 
the piezoelectric constant is 
zero\cite{Tsen70,BrGi79,AlCh81,Vela84,KoSy86,BrHu89}
or to the case where the free surface of the substrate is 
metallized\cite{BrGi79},
or in the weak piezoelectric coupling 
approximation\cite{BrGi79, KeSh76}. 
Here the crystal is cut along a plane containing the $Z$ axis and 
making \textit{any} angle with the $XY$ crystallographic plane. 
Moreover the surface may be metallized, or in contact with the 
vacuum, or in contact with a thin conducting layer with arbitrary 
finite conductance. 
Also, no approximation is made about the strength of the 
piezoelectric effect.
Attention is however limited to crystals with
tetragonal $\bar{4}$ symmetry (inclusive of the 
tetragonal $\bar{4}2$m, cubic $\bar{4}3$m, and cubic 23 symmetries). 
This limitation is not essential but simplifies the notation to a 
certain extent; 
it is nevertheless possible to extrapolate the method presented 
hereafter to crystals with lower symmetries such as orthorhombic 222, 
as pointed out at the end of the paper.
Note that the study of piezoacoustic SH surface waves in 
Potassium Niobate (2mm symmetry) was recently undertaken by 
Mozhaev and Weihnacht\cite{MoWe02} who showed that for that class,
the secular equation is a cubic in the squared wave speed.

The constitutive equations and the piezoacoustic equations for the 
class of crystals listed above are recalled in II.A and II.B, 
respectively, and some fundamental equations, which encapsulate the 
whole boundary value problem and its resolution, are quickly derived 
in II.C. 

These fundamental equations are applied in Section III to the 
consideration of a piezoacoustic SH surface wave. 
A method based on the resolution of the propagation condition for 
partial modes first, and the resolution of the boundary value 
conditions next, would lead to 
quite an involved analysis, including the analytical examination of a 
quartic polynomial for the coefficients of attenuation. 
The method based on the fundamental equations derived in II.C 
circumvents the stage of the quartic and delivers directly the secular 
equation in polynomial form. 
In particular it is seen that for metallized (a.k.a. short-circuit) 
boundary conditions (III.A), this equation is just a quadratic in the 
squared wave speed, whose relevant root is readily identified.
For open-circuit boundary conditions (III.B), the secular equation is 
also a quadratic in the squared wave speed, but it is not valid 
(the corresponding solution does not satisfy the boundary conditions.)
For the free (non-metallized) boundary condition (III.C), the secular 
equation is a sextic in the squared wave speed. 
For the conductive thin layer boundary condition (III.D),  the secular 
equation is a polynomial of degree 16 in the wave speed.  

Once the secular equation is solved for the speed, the complete 
description of the wave follows naturally (IV.A), including the 
attenuation coefficients and the profiles for the mechanical 
displacement, electric potential, traction, and electrical induction.
A simple check for the validity of the solution is proposed in 
IV.B.

The results are illustrated numerically and graphically using 
experimental data available for GaAs (cubic $\bar{4}3$m), 
BaLaGa$_3$O$_7$ (tetragonal $\bar{4}2$m) and Bi$_{12}$GeO$_{20}$ 
(cubic 23) in V. 
The range of existence of the free SH wave with respect to the angle 
of cut, the speeds of propagation, the amplitude of the profiles,  
etc. are all quantities which can be obtained numerically 
with as high a degree of numerical accuracy as is needed. 

The paper concludes with a discussion of the merits and possible 
applications of the method presented,
and with a chronological account of the several advances in the field, 
without which the results of this paper could not have been 
established. 

\section{Preliminaries} 

\subsection{Constitutive equations}

Consider a piezoelectric crystal with mass density $\rho$, 
possessing at most the tetragonal $\bar{4}$ symmetry
(this symmetry includes the tetragonal $\bar{4}2$m,  
cubic $\bar{4}3$m, and cubic $23$ cases.)
Let $\hat{c}_{ijkl}$, $\hat{e}_{ijk}$, and $\hat{\epsilon}_{ij}$ be 
its respective elastic, piezoelectric, and dielectric constants with 
respect to the coordinate system $XYZ$ along the crystallographic 
axes. 
Now cut the crystal by a plane containing the $Z$ axis and making 
an angle $\theta$ with the $XY$ plane. 
The new coordinate system $x_1 x_2 x_3$ (say), obtained after rotation 
of $XYZ$ about $Z$, is defined by 
\begin{equation} \label{rotation}
\begin{bmatrix}  x_1 \\ x_2 \\ x_3 \end{bmatrix} 
 = \begin{bmatrix}  \cos \theta & \sin \theta & 0 \\ 
                              -\sin \theta & \cos \theta & 0 \\
                               0 &  0 & 1 \end{bmatrix} 
\begin{bmatrix}  X \\ Y \\ Z  \end{bmatrix}, 
\end{equation} 
and so, the plane of cut is defined by $x_2 = 0$, as shown on 
Figure 1.

\begin{figure}
\begin{centering}
\epsfig{figure=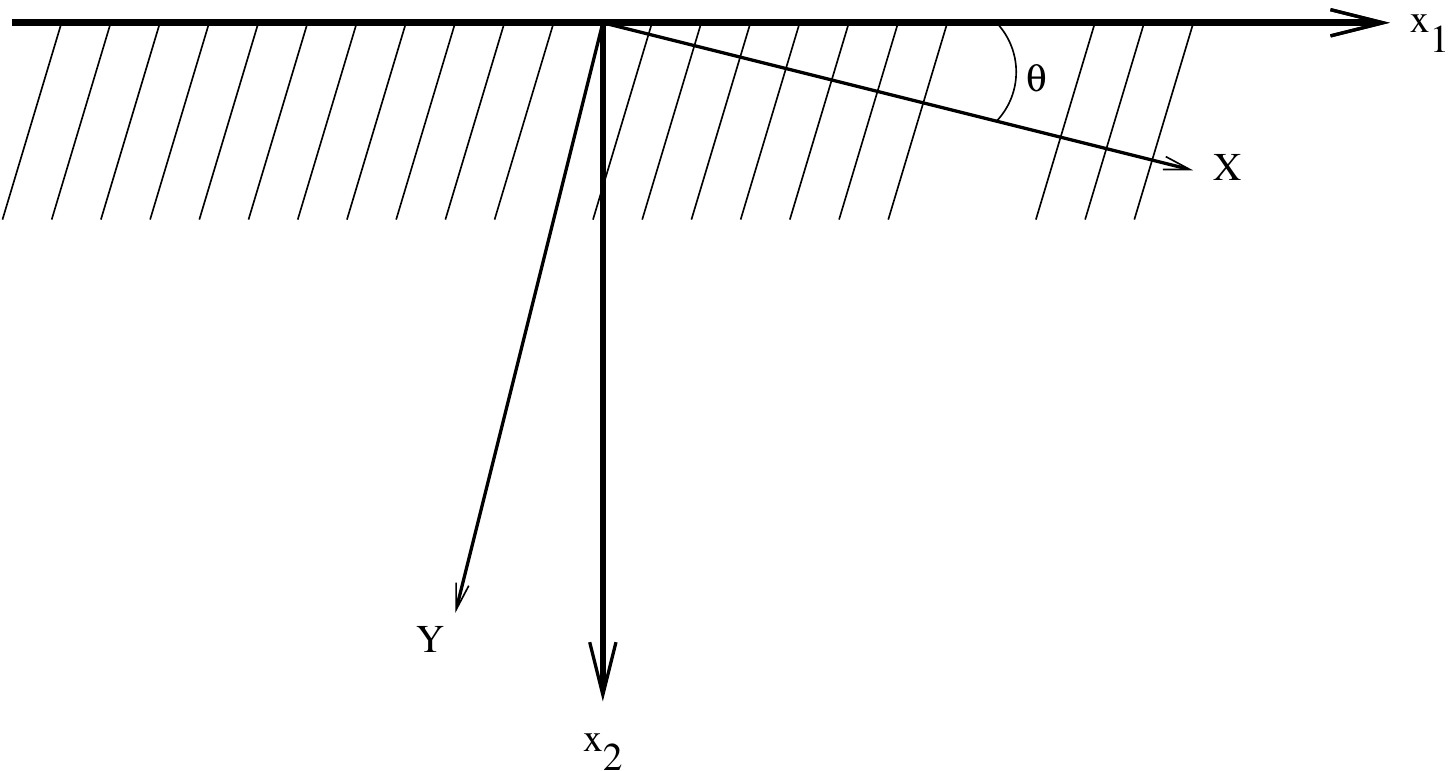, width=0.6\textwidth}
\caption{Rotated $Y$-cut about the $Z$ axis.}
\end{centering}
\end{figure}

In the new coordinate system, under the electrostatic approximation 
for the electrical field,  the stress tensor components $\sigma_{ij}$ 
and the electric induction components $D_i$ are related to the 
gradients of the mechanical displacement $\mathbf{u}$ and of the 
electrical potential $\phi$ by the constitutive relations,
\begin{equation} \label{constitutive}
 \sigma_{ij} = c_{ijkl} u_{l,k} + e_{ijk} \phi_{,k}, 
\quad 
 D_i = e_{ikl} u_{l,k} - \epsilon_{ik} \phi_{, k}, 
\end{equation}
where the comma denotes partial differentiation (here with respect to 
the $x_k$ coordinates). 
Using the Voigt contracted notation for the $c$, $e$, and $\epsilon$, 
these relations are written in  matrix form as 
\begin{equation} \label{StressStrainGeneral}
\begin{bmatrix} 
   \sigma_{11} \\
   \sigma_{22} \\
   \sigma_{33} \\
   \sigma_{23} \\
   \sigma_{31} \\
   \sigma_{12} \\
   D_1 \\
   D_2 \\
   D_3 
\end{bmatrix}
=
\begin{bmatrix}
c_{11} & c_{12} & c_{13} & 0 & 0 & c_{16} & 0 & 0 & e_{31} \\
c_{12} & c_{11} & c_{13} & 0 & 0 & -c_{16} & 0 & 0 & -e_{31} \\
c_{13} & c_{13} & c_{33} & 0 & 0 & 0 & 0 & 0 & 0         \\
0 & 0 & 0 & c_{44} & 0 & 0 & e_{14} & -e_{15} & 0  \\
0 & 0 & 0 & 0 & c_{44} & 0 & e_{15} & e_{14} & 0    \\
c_{16} & -c_{16} & 0 & 0 & 0 & c_{66} & 0 & 0 & e_{36} \\
0 & 0 & 0 & e_{14} & e_{15} & 0 & -\epsilon_{11} & 0 & 0 \\
0 & 0 & 0 & -e_{15} & e_{14} & 0 & 0 & -\epsilon_{11} & 0 \\
e_{31} & -e_{31} & 0 & 0 & 0 & e_{36} & 0 & 0 & -\epsilon_{33}
 \end{bmatrix}
\begin{bmatrix} 
   u_{1,1} \\
   u_{2,2} \\
   u_{3,3} \\
   u_{2,3} + u_{3,2} \\
   u_{3,1} + u_{1,3} \\
   u_{1,2} + u_{2,1} \\
   \phi_{,1} \\ 
   \phi_{,2} \\ 
   \phi_{,3} \\ 
\end{bmatrix}.
\end{equation}

Explicitly, 
the $c_{ij}$ and $e_{ij}$ are deduced from the $\hat{c}_{ij}$ and 
$\hat{e}_{ij}$ in $XYZ$ by well-known relationships. 
In particular,
\begin{align} \label{rotate}
& c_{44} = \hat{c}_{44}, \quad 
\epsilon_{11} = \hat{\epsilon}_{11}, 
\nonumber \\ 
& e_{14} = \hat{e}_{14} \cos 2\theta  - \hat{e}_{15} \sin 2\theta, 
\nonumber \\ 
& e_{15} = \hat{e}_{15} \cos 2\theta  + \hat{e}_{14} \sin 2\theta.
\end{align}

\subsection{Piezoacoustic equations}

Now consider the propagation of an anti-plane (SH) inhomogeneous wave 
in the half-space $x_2 \ge 0$, travelling with speed $v$ and wave 
number $k$ in the $x_1$ direction, with attenuation in the $x_2$ 
direction. 
It is known\cite{FaAd72} that for the crystals under consideration 
this wave decouples entirely from its in-plane counterpart, a purely 
elastic two-component Rayleigh wave. 
Thus the wave is modelled as $u_1 = u_2 = 0$, and  
\begin{equation} \label{wave} 
\{ u_3, \phi \} =
 \{ U_3(kx_2), \varphi(kx_2) \}e^{ik (x_1 - vt)},
\end{equation}
for some yet unknown functions $U_3$ and $\varphi$ of $kx_2$.
Accordingly, the constitutive equations Eq.~\eqref{constitutive} 
lead to similar forms for the stress and electrical components, 
\begin{equation} \label{wave2} 
\{ \sigma_{ij}, D_i \} =
 ik \{ t_{ij}(kx_2), d_i(kx_2) \}e^{ik (x_1 - vt)},
\end{equation}
where $t_{11} = t_{22} = t_{12} = d_3 = 0$, and 
\begin{align} 
&t_{33} = c_{13} U_1 - i c_{11} U'_2, 
\nonumber \\ 
& t_{32} = -ic_{44} U'_3 + e_{14} \varphi  + ie_{15} \varphi', 
\nonumber \\ 
&  t_{31} = c_{44} U_3 + e_{15} \varphi  - ie_{14} \varphi', 
\nonumber \\ 
& d_1 = e_{15} U_3 - ie_{14} U'_3  - \epsilon_{11} \varphi, 
\nonumber \\ 
& d_2 = e_{14} U_3 + ie_{15} U'_3  + i\epsilon_{11} \varphi'. 
\end{align}
Here and hereafter, the prime denotes differentiation with respect to 
the variable $kx_2$.

Using the above-introduced functions $U_3$, $\varphi$, $t_{ij}$, 
 $d_i$, the classical equations of piezoacoustics, 
\begin{equation} 
\sigma_{ij,j} =  \rho u_{i,tt}, 
 \quad D_{i,i} = 0, 
\end{equation} 
reduce to 
\begin{equation} \label{piezoacoustic1} 
 - t_{13} + i t'_{32} = -X U_3, \quad
  - d_1 + i d'_2 = 0, 
\end{equation} 
where $X := \rho v^2$.  

When the region $x_2 < 0$ above the crystal is the vacuum 
(permeability: $\epsilon_0$) and the 
surface $x_2 = 0$ remains free of tractions, then the boundary value 
problem corresponding to Eq.~\eqref{piezoacoustic1} is that of 
piezoacoustic (Bleustein-Gulyaev) SH surface waves.

\subsection{Fundamental equations for the resolution}

The method of resolution rests on the property that the piezoacoustic 
equations \eqref{piezoacoustic1} 
can be written in the form,
\begin{equation}  \label{motion}
\mbox{\boldmath $\xi$}'= i \mathbf{N}\mbox{\boldmath $\xi$}, 
\quad \text{where} \quad 
\mbox{\boldmath $\xi$}(kx_2) = 
 [U_3, \varphi, t_{32}, d_2]^\text{T}, 
\end{equation}
and $\mathbf{N}$ is a real 4$\times$4 matrix which, 
brought up to any positive or negative integer power $n$ has the 
following block structure, 
\begin{equation} \label{structure}
\mathbf{N}^n = \begin{bmatrix}
 \mathbf{N^{(n)}_1} & \mathbf{N^{(n)}_2} \\
 \mathbf{K^{(n)}}
        &  (\mathbf{N^{(n)}_1})^{\mathrm{T}}
\end{bmatrix}, \quad 
\text{with} \quad
\mathbf{K^{(n)}} = (\mathbf{K^{(n)}})^{\mathrm{T}}, 
\quad
\mathbf{N^{(n)}_2} = (\mathbf{N^{(n)}_2})^{\mathrm{T}},
\end{equation}
where $^\text{T}$ denotes the transpose, and where the submatrices 
of $\mathbf{N}^n$ are $2 \times 2$ matrices. 
The first-order differential form 
Eq.~\eqref{motion} of the piezoacoustic equations dates back to 
Kraut\cite{Krau69}, and before that, to Stroh\cite{Stro58} and 
others (see Fahmy and Adler\cite{FaAd73} for references) for the 
purely elastic case.
The subsequent analysis below builds upon several crucial 
contributions, which are listed and put into context at the very end 
of this article.

Now, because the wave amplitude must vanish away from 
$x_2=0$, $\mbox{\boldmath $\xi$}(kx_2)$ is such that 
\begin{equation} \label{infinity}
\mbox{\boldmath $\xi$}(\infty) = \mathbf{0}.
\end{equation}

Clearly, pre-multiplication of $\mathbf{N}^n$ by 
$\widehat{\mathbf{I}}$ defined as,
\begin{equation} 
 \widehat{\mathbf{I}} = 
  \begin{bmatrix} 0 & \mathbf{1} \\
                           \mathbf{1} & 0 
  \end{bmatrix}, \quad \text{where} \quad 
\mathbf{1}  = 
  \begin{bmatrix} 1 & 0 \\
                           0  & 1
  \end{bmatrix},
\end{equation}
produces a symmetric matrix, 
\begin{equation} 
   \widehat{\mathbf{I}}\mathbf{N}^n =
\begin{bmatrix}
 \mathbf{K^{(n)}} & (\mathbf{N^{(n)}_1})^{\mathrm{T}} \\
  \mathbf{N^{(n)}_1} & \mathbf{N^{(n)}_2}
\end{bmatrix}
= (\widehat{\mathbf{I}}\mathbf{N}^n)^\text{T}, 
\end{equation}
so that taking the scalar product on both sides of Eq.~\eqref{motion}
by 
$\widehat{\mathbf{I}}\mathbf{N}^n \overline{\mbox{\boldmath $\xi$}}$,
where the overbar denotes the complex conjugate, leads to 
$\overline{\mbox{\boldmath $\xi$}} \cdot 
 \widehat{\mathbf{I}}\mathbf{N}^n \mbox{\boldmath $\xi$}' = 
i \overline{\mbox{\boldmath $\xi$}} \cdot 
 \widehat{\mathbf{I}}\mathbf{N}^{n+1} \mbox{\boldmath $\xi$}$, 
the right hand-side of which is purely imaginary. 
Taking the real part and integrating yields 
$\overline{\mbox{\boldmath $\xi$}} \cdot 
 \widehat{\mathbf{I}}\mathbf{N}^n \mbox{\boldmath $\xi$} = $
const. $=0$ (by Eq.~\eqref{infinity}), and in particular,
\begin{equation} \label{equations}
\overline{\mbox{\boldmath $\xi$}}(0)  \cdot 
   \widehat{\mathbf{I}}\mathbf{N}^n \mbox{\boldmath $\xi$}(0) =  0.
\end{equation}
These \textit{fundamental equations}\cite{Dest03d, Dest04a, Dest04b} 
allow for a completely analytical derivation of the secular equation, 
for a great variety of boundary conditions. 
Note that because $\mathbf{N}$ is a $4 \times 4$ matrix, there are 
at most three independent fundamental equations according to the 
Cayley-Hamilton theorem. 
Any choice of three different integers $n$ is legitimate, although 
the choice $n=-1,1,2$ seems to yield the most compact expressions 
for the components of $\mathbf{N}^n$.

\section{Secular equations}

For the piezoacoustic (Bleustein-Gulyaev) shear-horizontal wave, 
the matrix $\mathbf{N}$ in Eq.~\eqref{motion}
is written in compact form using the following quantity,
\begin{equation} \label{K}
\kappa^2 = \dfrac{e^2_{14}}{\epsilon_{11}c_{44} + e_{15}^2} , 
\end{equation} 
as 
\begin{equation} \label{NforSH}
\mathbf{N} = 
\begin{bmatrix} 
\vspace{6pt}
\dfrac{e_{15}}{e_{14}}\kappa^2 
  & -\dfrac{\epsilon_{11}}{e_{14}}\kappa^2 
& \dfrac{\epsilon_{11}}{e_{14}^2}\kappa^2 
  & -\dfrac{e_{15}}{e_{14}^2}\kappa^2 \\
\vspace{6pt}
\dfrac{c_{44}}{e_{14}}\kappa^2 & \dfrac{e_{15}}{e_{14}}\kappa^2 
& -\dfrac{e_{15}}{e_{14}^2}\kappa^2 
  & -\dfrac{c_{44}}{e_{14}^2}\kappa^2 \\
\vspace{6pt}
X - c_{44}(1+\kappa^2) & -e_{15}(1+\kappa^2) 
& \dfrac{e_{15}}{e_{14}}\kappa^2 & \dfrac{c_{44}}{e_{14}}\kappa^2 \\
\vspace{6pt}
- e_{15}(1+\kappa^2) & \epsilon_{11}(1+\kappa^2) 
& -\dfrac{\epsilon_{11}}{e_{14}}\kappa^2 
  & \dfrac{e_{15}}{e_{14}}\kappa^2 
\end{bmatrix}. 
\end{equation} 

Note that $\mathbf{N}$ is indeed of the form Eq.~\eqref{structure}.
To express explicitly integer powers of $\mathbf{N}$, it proves 
convenient to use scalar multiples of the matrix 
$\widehat{\mathbf{I}}\mathbf{N}^n$, for which the fundamental 
equations Eq.~\eqref{equations} are also valid.
For instance, the fundamental equations \eqref{equations} 
also hold at $n=2,-1$ when 
$\widehat{\mathbf{I}}\mathbf{N}^n$ is replaced with $\mathbf{M^{(2)}}$,
$\mathbf{M^{(-1)}}$ respectively, defined by
\begin{equation}
\mathbf{M^{(2)}} = 
 \dfrac{e_{14}}{e_{15} \kappa^2}
                                  \widehat{\mathbf{I}}\mathbf{N}^2,
\quad 
\mathbf{M^{(-1)}} = 
 (\dfrac{\epsilon_{11}c_{44}}{e_{14}^2}\kappa^2 X - 1)
                                  \widehat{\mathbf{I}}\mathbf{N}^{-1}. 
\end{equation} 
These symmetric matrices are given explicitly by 
their components,
\begin{align} \label{M(2)}
&     M^{(2)}_{11} = 2[X- 2c_{44}(1+\kappa^2)],
&&   M^{(2)}_{12} = 
   -\dfrac{\epsilon_{11}}{e_{15}} X 
       + 2\dfrac{\epsilon_{11}c_{44}-e_{15}^2}{e_{15}}(1+\kappa^2), 
\nonumber \\
&     M^{(2)}_{22} = 4\epsilon_{11}(1+\kappa^2),
&&   M^{(2)}_{13} = 
        \dfrac{\epsilon_{11}}{e_{14}e_{15}} X 
    - \dfrac{\epsilon_{11}c_{44}-e_{15}^2}{e_{14}e_{15}}(1+2\kappa^2),
\nonumber \\
&     M^{(2)}_{23} = 
         -2\dfrac{\epsilon_{11}}{e_{14}}(1+2\kappa^2),
&&   M^{(2)}_{33} = 
         4\dfrac{\epsilon_{11}}{e_{14}^2}\kappa^2, 
\nonumber \\
&    M^{(2)}_{14} = 
         -\dfrac{1}{e_{14}}X + 2\dfrac{c_{44}}{e_{14}}(1+2\kappa^2),
&&  M^{(2)}_{24} = 
      -\dfrac{\epsilon_{11}c_{44}-e_{15}^2}{e_{14}e_{15}}(1+2\kappa^2),
\nonumber \\
&   M^{(2)}_{34} = 
      2\dfrac{\epsilon_{11}c_{44}-e_{15}^2}{e_{14}^2 e_{15}}\kappa^2, 
&& M^{(2)}_{44} = 
       -4\dfrac{c_{44}}{e_{14}^2}\kappa^2,
\end{align}    
and
\begin{align} \label{M(-1)}
&     M^{(-1)}_{11} = 
   (1+\dfrac{\epsilon_{11}c_{44}}
       {e_{14}^2})\kappa^2 X - c_{44}(1+\kappa^2),
&&   M^{(-1)}_{12} = 
   -\dfrac{\epsilon_{11}e_{15}}{e_{14}^2}\kappa^2 X 
                      + e_{15}(1+\kappa^2), 
\nonumber \\
&     M^{(-1)}_{22} = 
   -\dfrac{\epsilon_{11}^2}
       {e_{14}^2}\kappa^2 X + \epsilon_{11}(1+\kappa^2),
&&   M^{(-1)}_{13} = 
         \dfrac{e_{15}}{e_{14}}\kappa^2, 
\nonumber \\
&     M^{(-1)}_{23} = 
         \dfrac{\epsilon_{11}}{e_{14}}\kappa^2,
&&   M^{(-1)}_{33} = 
         \dfrac{\epsilon_{11}}{e_{14}^2}\kappa^2, 
\nonumber \\
&    M^{(-1)}_{14} = 
         \dfrac{X-c_{44}}{e_{14}}\kappa^2,
&&  M^{(-1)}_{24} = 
         \dfrac{e_{15}}{e_{14}}\kappa^2, 
\nonumber \\
&   M^{(-1)}_{34} = 
         \dfrac{e_{15}}{e_{14}^2}\kappa^2, 
&& M^{(-1)}_{44} = 
       \dfrac{X-c_{44}}{e_{14}^2}\kappa^2.
\end{align}    

Now all the required equations and quantities are in place to treat 
various electrical boundary value problems.

\subsection{Metallized (short-circuit) boundary condition}

Here the surface of the crystal is coated with a thin metallic film, 
with thickness negligible when compared to the wavelength, and brought 
to a zero electrical potential. 
Moreover, the coating still allows the surface to remain free of 
mechanical tractions. 
Then 
\begin{equation}
\sigma_{23} = 0, \: \phi = 0, \text{ at } x_2 = 0, \quad 
\text{so that} \quad
\mbox{\boldmath $\xi$}(0) = [U_3(0), 0, 0, d_2(0)]^\text{T}.
\end{equation}
Writing $\mbox{\boldmath $\xi$}(0)= U_3(0)[1, 0, 0, \alpha]$, where 
$\alpha = d_2(0)/U_3(0)$ is complex, the fundamental equations 
\eqref{equations} for 
$\widehat{\mathbf{I}}\mathbf{N}$ ($n=1$), 
$\mathbf{M^{(2)}}$ ($n=2$), and $\mathbf{M^{(-1)}}$ ($n=-1$) 
lead to the following homogeneous system of equations,
\begin{equation} \label{secul2}
\begin{bmatrix} 
   N_{31}          &   N_{21}           &   N_{24}  \\
   M^{(2)}_{11}  &   M^{(2)}_{14}   &  M^{(2)}_{44}  \\
   M^{(-1)}_{11} &   M^{(-1)}_{14}  &  M^{(-1)}_{44} 
\end{bmatrix}
\begin{bmatrix} 
 1 \\ \alpha +\overline{\alpha} \\ \alpha\overline{\alpha}
\end{bmatrix} 
  = \begin{bmatrix} 
     0 \\ 0 \\ 0
     \end{bmatrix}.
\end{equation}
For a non-trivial solution to exist, the determinant of the system's 
matrix must be zero. 
Factoring out common factors, this condition reads 
\begin{equation}
\begin{vmatrix}
X-c_{44}(1+\kappa^2) & c_{44}\kappa^2 & -c_{44} \\
2[X-2c_{44}(1+\kappa^2)] & -X+2c_{44}(1+2\kappa^2) & -4c_{44} \\
(1+\dfrac{\epsilon_{11}c_{44}}{e_{14}^2})\kappa^2 X-c_{44}(1+\kappa^2)
 & (X-c_{44})\kappa^2 & X-c_{44}
\end{vmatrix} = 0.
\end{equation}
When $X=0$, the first column in the determinant becomes proportional 
to the third, and the determinant is zero. 
Hence $X$ is a factor of the determinant;
the remaining factor is a quadratic in $X$,
\begin{equation} \label{secularShort}
X^2 
- c_{44}(\dfrac{3\epsilon_{11}c_{44}+4e_{14}^2 +4e_{15}^2}
                                  {\epsilon_{11}c_{44}+e_{15}^2}) X
+ 2c_{44}^2(\dfrac{\epsilon_{11}c_{44}+2e_{14}^2+2e_{15}^2}
                                  {\epsilon_{11}c_{44}+e_{15}^2}) 
=0,
\end{equation}
that is, the \textit{explicit secular equation for 
piezoacoustic (Bleustein-Gulyaev) 
anti-plane surface waves on a metallized 
tetragonal $\bar{4}$ (or tetragonal $\bar{4}2$m, or cubic $\bar{4}3$m, 
$23$) crystal, cut along any plane containing the $Z$ axis}.

This equation being a quadratic in $X$, it is solved explicitly 
and it yields \textit{a priori} two roots. 
The selection is made by considering the known speed of a 
Bleustein-Gulyaev surface wave in a cubic $\bar{4}3$m or $23$ 
crystal when $\theta = 45^\text{o}$. 
Then the root of the quadratic corresponding to the plus sign  
is $X=\rho v^2 = 2 c_{44}$, and the root 
corresponding to the minus sign is 
\begin{equation} \label{knownSpeed}
\rho v^2 = 
 c_{44}(1+\dfrac{e_{15}^2}{c_{44}\epsilon_{11}+e_{15}^2}) = 
  c_{44}(1 + 
      \dfrac{\hat{e}_{14}^2}{c_{44}\epsilon_{11}+\hat{e}_{14}^2}), 
\end{equation}
in accordance with Tseng\cite{Tsen70} (see also 
Koerber and Vogel\cite{KoVo72}, Alburque and Chao\cite{AlCh81}, 
Velasco\cite{Vela84}, Bright and Hunt\cite{BrHu89}).
By continuity with the other cases ($\bar{4}$, $\bar{4}2$m, 
$\bar{4}3$m, $23$, $\theta \ne  45^\text{o}$), the 
\textit{speed of the generic Bleustein-Gulyaev wave} found from the 
secular equation \eqref{secularShort} is $v_\text{BGm}$, given by 
\begin{equation} \label{speedShort}
\dfrac{\rho v_\text{BGm}^2}{c_{44}} = 
  \dfrac{3\epsilon_{11}c_{44} + 4 e_{14}^2 + 4e_{15}^2 -
       [(\epsilon_{11}c_{44} + 4e_{14}^2)^2 
              + (4e_{14}e_{15})^2]^{\textstyle{\frac{1}{2}}}}
             {2(\epsilon_{11}c_{44} + e_{15}^2)}.
\end{equation}
For tetragonal $\bar{4}2$m, cubic $\bar{4}3$m, 
or cubic $23$ crystals, $\hat{e}_{15} =0$ and by Eq.~\eqref{rotate}, 
this expression reduces to 
\begin{equation} \label{speedShortCubic}
\dfrac{\rho v_\text{BGm}^2}{c_{44}} = 
  \dfrac{3 + 4 \hat{\chi}^2 -
       [(1 + 4\hat{\chi}^2 \cos^2 2\theta)^2 
           + 4\hat{\chi}^4 \sin^2 4\theta]^{\textstyle{\frac{1}{2}}}}
             {2(1 + \hat{\chi}^2 \sin^2 2\theta)},
\end{equation}
as proved by Braginski\u{i} and Gilinski\u{i}\cite{BrGi79} using a 
different method. 
In Eq.~\eqref{speedShortCubic}, 
$\hat{\chi}^2 = \hat{e}_{14}^2 / (\epsilon_{11}c_{44})$ is the 
``piezoelectric coupling coefficient'' for bulk waves.

\subsection{Electrically open boundary condition}

The substrate is said to be ``mechanically free, electrically open'' 
(Ingebrigsten\cite{Inge69}, Lothe and Barnett\cite{BaLo76}) when 
\begin{equation}
\sigma_{32} = 0, \; D_2 = 0, \text{ at } x_2 = 0, \quad 
\text{so that} \quad
\mbox{\boldmath $\xi$}(0) = [U_3(0), \varphi(0), 0, 0]^\text{T}.
\end{equation}

Writing $\mbox{\boldmath $\xi$}(0)= U_3(0)[1,  \alpha, 0, 0]$, where 
$\alpha = \varphi(0)/U_3(0)$ is complex, the fundamental equations 
\eqref{equations} for 
$\widehat{\mathbf{I}}\mathbf{N}$ ($n=1$), 
$\mathbf{M^{(2)}}$ ($n=2$), and $\mathbf{M^{(-1)}}$ ($n=-1$) 
lead to the following homogeneous system of equations,
\begin{equation} \label{secul3}
\begin{bmatrix} 
   N_{31}        &   N_{32}         &  N_{42}  \\
   M^{(2)}_{11}  &   M^{(2)}_{12}   &  M^{(2)}_{22}  \\
   M^{(-1)}_{11} &   M^{(-1)}_{12}  &  M^{(-1)}_{22} 
\end{bmatrix}
\begin{bmatrix} 
 1 \\ \alpha +\overline{\alpha} \\ \alpha\overline{\alpha}
\end{bmatrix} 
  = \begin{bmatrix} 
     0 \\ 0 \\ 0
     \end{bmatrix}.
\end{equation}
For a non-trivial solution to exist, the determinant of the system's 
matrix must be zero. 
Its components are  given in Eq.~\eqref{NforSH},  Eq.~\eqref{M(2)}, 
and Eq.~\eqref{M(-1)}.
As in the short-circuit configuration, at $X=0$ the first and third 
columns in the determinant become both proportional to 
$[1,4,1]^\text{T}$ and so $X$ is a factor of the determinant;
the remaining factor is a quadratic in $X$,
\begin{equation} \label{secularOpen}
\dfrac{\epsilon_{11}^2}{e_{14}^2} \kappa^2 X^2 
- \epsilon_{11}(1+ \kappa^2)
       [3 - (1+4\dfrac{e_{15}^2}{e_{14}^2})\kappa^2]X 
+ 2(1+ \kappa^2)^2(\epsilon_{11}c_{44}-e_{14}^2-e_{15}^2) 
=0.
\end{equation}

At this stage, an important point must be raised.
Although this secular equation might seem legitimate at first sight, 
it must be recalled that it was obtained through a process based on 
the fundamental equations \eqref{equations} 
which, due to the involvement of integer powers 
of the matrix $\mathbf{N}$, might generate spurious secular equations.
In fact, it has been proved\cite{BaLo76, BrGi79} 
that the Bleustein-Gulyaev wave \textit{does not exist for open
circuit boundary conditions}. 
Hence the secular equation Eq.~\eqref{secularOpen} is not valid. 
It is given here for completeness and to illustrate one limitation 
of the fundamental equations approach. 
However, as is seen in IV.B, a simple check can be done 
to realize whether the speed given by an explicit secular equation is 
valid or not.

\subsection{Free boundary condition}

In the general case of a non-metallized, mechanically free boundary, 
the tangential component of the electric field and the normal 
component of the electric induction are continuous across the 
substrate/vacuum interface. 
These continuities lead to the relationship (e.g.~Dieulesaint and 
Royer \cite{DiRo80} p.288), 
\begin{equation}
  d_2(0) = i \epsilon_0 \varphi(0), 
  \quad \text{so that} \quad
   \mbox{\boldmath $\xi$}(0)
     = \varphi(0)[\alpha, 1, 0, i\epsilon_o]^\text{T},   
\end{equation}
where $\alpha =\alpha_1 + i\alpha_2 = U_3(0)/\varphi(0)$ is complex. 
Now the fundamental equations \eqref{equations} for 
$\widehat{\mathbf{I}}\mathbf{N}$ ($n=1$), 
$\mathbf{M^{(2)}}$ ($n=2$), and $\mathbf{M^{(-1)}}$ ($n=-1$) 
lead to a non-homogeneous linear system of equations,
\begin{equation} \label{system4}
\begin{bmatrix} 
   N_{32}        & \epsilon_0 N_{21}         &  N_{31}        \\
   M^{(2)}_{12}  & \epsilon_0 M^{(2)}_{14}   &  M^{(2)}_{11}  \\
   M^{(-1)}_{12} & \epsilon_0 M^{(-1)}_{14}  &  M^{(-1)}_{11} 
\end{bmatrix}
\begin{bmatrix} 
 2\alpha_1  \\ 2 \alpha_2 \\ \alpha_1^2 + \alpha_2^2
\end{bmatrix} 
  = \begin{bmatrix} 
     - N_{42}        - \epsilon_0^2 N_{24}       \\ 
     - M^{(2)}_{22}  - \epsilon_0^2 M^{(2)}_{44} \\ 
     - M^{(-1)}_{22} - \epsilon_0^2 M^{(-1)}_{44}  
     \end{bmatrix}.
\end{equation}
By Cramer's rule, the unique solution to this system is 
\begin{equation} \label{Cramer}
  2\alpha_1 = \Delta_1/\Delta, \quad 
  2\alpha_2 = \Delta_2/\Delta, \quad 
  \alpha_1^2 + \alpha_2^2 = \Delta_3/\Delta,
\end{equation}
where $\Delta$ is the determinant of the $3 \times 3$ matrix 
in Eq.~\eqref{system4}, with components given in Eq.~\eqref{NforSH}, 
Eq.~\eqref{M(2)}, and Eq.~\eqref{M(-1)}, and the $\Delta_k$ are the 
determinants obtained by replacing this matrix's $k$-th column 
with the vector on the right hand-side of Eq.~\eqref{system4}.
It follows from Eq.~\eqref{Cramer} that
\begin{equation} \label{seculFree}
  \Delta_1^2 + \Delta_2^2 - 4 \Delta \Delta_3 =0,
\end{equation}
which is the \textit{explicit secular equation for 
piezoacoustic (Bleustein-Gulyaev) anti-plane surface waves on a 
non-metallized, mechanically free $\bar{4}$ 
(or $\bar{4}2$m, $\bar{4}3$m, 
 $23$) crystal, cut along any plane containing the $Z$ axis}.

The expansions of the determinants $\Delta$, \ldots, $\Delta_3$ are 
lengthy and are not displayed here, but they are easily computed by 
using the components  Eq.~\eqref{NforSH}, Eq.~\eqref{M(2)}, 
and Eq.~\eqref{M(-1)}.
It turns out that $\Delta$ factorizes into the product of $\epsilon_0$
and a cubic in $X$ which is independent of $\epsilon_0$, while 
$\Delta_1$ (resp. $\Delta_2$, $\Delta_3$) factorizes into the 
product of $\epsilon_0 X$ (resp. $X$, $\epsilon_0$) and a polynomial 
which is quadratic in $X$ and linear in $\epsilon_0^2$.  
The resulting secular equation \eqref{seculFree} is a sextic in 
$X$ and a cubic in $\epsilon_0^2$, with the coefficient of the 
$\epsilon_0^6$ term proportional to the 
``metallized secular equation'' \eqref{secularShort} and the 
coefficient of the $\epsilon_0^0$ term proportional to the (non-valid) 
``open circuit secular equation'' \eqref{secularOpen}. 

It is emphasized again that, as in III.B, great 
care must be taken to ensure that the speed given by the secular 
equation Eq.~\eqref{seculFree} leads to a valid solution. 
This point is discussed in IV.B.

\subsection{Thin conducting layer boundary condition}

As a final type of boundary condition, consider that the semi-infinite 
substrate is covered with a metallic film with thickness $h$ 
and conductance $\gamma$, where $h$ is assumed to be so small 
with respect to the acoustic wavelength that the effects of mechanical 
loading can be neglected.
Then the permeability of the region $x_2<0$ close to the interface is 
changed from $\epsilon_0$ (see previous Subsection) to 
$\epsilon_0 - i \epsilon$ (see Royer and Dieulesaint\cite{RoDi96}, 
p. 301), with $\epsilon = \gamma h / v$. 
Replacing the former quantity with the latter in the previous 
Subsection leads to similar results as above, except that the 
$3 \times 3$ matrix and the right hand-side of Eq.~\eqref{system4} 
are now replaced with 
\begin{multline} \label{system5}
\begin{bmatrix} 
   N_{32} + \epsilon N_{21} & \epsilon_0 N_{21} &  N_{31}        \\
   M^{(2)}_{12} + \epsilon M^{(2)}_{14}  & \epsilon_0 M^{(2)}_{14}   
       &  M^{(2)}_{11}  \\
   M^{(-1)}_{12}  + \epsilon M^{(-1)}_{14} 
       & \epsilon_0 M^{(-1)}_{14}   &  M^{(-1)}_{11} 
\end{bmatrix},
\\ \text{and} \quad  
 \begin{bmatrix} 
       -N_{42} - 2 \epsilon N_{22} 
                - (\epsilon_0^2 + \epsilon^2) N_{24} \\ 
       -M^{(2)}_{22}   - 2 \epsilon M^{(2)}_{24}
          - (\epsilon_0^2 + \epsilon^2) M^{(2)}_{44} \\ 
       -M^{(-1)}_{22}  - 2 \epsilon M^{(-1)}_{24}
          - (\epsilon_0^2 + \epsilon^2)  M^{(-1)}_{44}  
     \end{bmatrix},
\end{multline}
respectively.
Then the secular equation is Eq.~\eqref{seculFree} where 
$\Delta$, \ldots, $\Delta_3$ are appropriately changed. 
Because here the quantity $\epsilon$ depends upon $v$, the secular 
equation is a polynomial in the speed of degree higher than in the 
previous Subsection, namely it is a polynomial of degree 16 in $v$.

\section{Construction of the solutions}

\subsection{Description of the wave}

Once a wave speed is determined from the secular equation, 
it is a rather straightforward matter to construct the corresponding 
complete solution. 
Indeed, the anti-plane mechanical displacement $u_3$, 
the electrical potential $\phi$, the shear stress $\sigma_{32}$, 
and the electric induction $D_2$ are given by 
\begin{equation}
[u_3, \phi, \sigma_{32}, D_2](x_1,x_2,t)  = 
 \Re \{ 
 [U_3,\varphi,ikt_{32},ikd_2](kx_2)e^{ik(x_1-vt)} 
       \}.
\end{equation}
Here $U_3$, $\varphi$, $t_{32}$, $d_2$ are the components of 
$\mbox{\boldmath $\xi$}$, solution to 
$\mbox{\boldmath $\xi$}' = i \mathbf{N} \mbox{\boldmath $\xi$}$.
Taking $\mbox{\boldmath $\xi$}$ in exponential form leads to the 
following decaying solution,
\begin{equation}
\mbox{\boldmath $\xi$} = 
 \beta_1 \mbox{\boldmath $\xi^1$}e^{ikq_1x_2} 
 + \beta_2 \mbox{\boldmath $\xi^2$}e^{ikq_2x_2},
\end{equation}
where $\beta_1$, $\beta_2$ are constants, $q_1$, $q_2$ are the 
two roots with positive imaginary part to the \textit{inhomogeneous 
wave propagation condition}: 
$\text{det } (\mathbf{N} - q\mathbf{1})=0$, 
and the $\mbox{\boldmath $\xi^i$}$ satisfy: 
$\mathbf{N} \mbox{\boldmath $\xi^i$} = q_i\mbox{\boldmath $\xi^i$}$.

Explicitly, the $q_i$ are roots of the quartic, 
\begin{multline} \label{inhomPropCond}
(\epsilon_{11}c_{44}+e_{15}^2)q^4
                - 4e_{14}e_{15} q^3 
       - [\epsilon_{11}(X-2c_{44}) - 4e_{14}^2 + 2e_{15}^2]q^2 \\
 +  4e_{14}e_{15} q 
   + e_{15}^2 - \epsilon_{11}(X-c_{44}) = 0,
\end{multline}
and the $\mbox{\boldmath $\xi^i$}$ are proportional to any column 
vector of the matrix adjoint to $\mathbf{N} - q_i\mathbf{1}$, the 
third one say. 
Hence,
\begin{equation}
\mbox{\boldmath $\xi^i$} = 
 \left[ \dfrac{e_{14}}{c_{44}}a_i, 
            \: b_i, 
              \: e_{14}^2 f_i, 
                \: \dfrac{e_{14}^2}{c_{44}}g_i \right]^\text{T},
\end{equation}
where the non-dimensional real quantities $a_i$, $b_i$, $f_i$, $g_i$ 
($i=1,2$) are given by
\begin{align}
& a_i = 
  -\dfrac{\epsilon_{11}c_{44}}{e_{14}^2}(q_i^2+1), 
\quad
b_i = \dfrac{e_{15}}{e_{14}}(q_i^2-1) - 2q_i,
\nonumber \\
& f_i=  \dfrac{e_{15}}{e_{14}}(3q_i^2-1) 
   - \dfrac{e_{15}^2}{e_{14}^2}q_i(q_i^2 -1)
    - \dfrac{\epsilon_{11}c_{44}}{e_{14}^2}q_i(q_i^2+1)
     - 2q_i,
\nonumber \\
& g_i = \dfrac{\epsilon_{11}c_{44}}{e_{14}^2}
         (q_i^2 + 2\dfrac{e_{15}}{e_{14}}q_i -1).
\end{align}
Finally, the ratio $\beta_2/\beta_1$ comes from the condition that the 
third component of $\mbox{\boldmath $\xi$}(0)$, proportional to 
$t_{23}(0)$, must be zero, so that $\beta_2 / \beta_1 = - f_1 / f_2$.

\subsection{Validity of the solution}

Once a wave solution has been constructed, its validity 
must be checked that is, it must satisfy the boundary conditions.

Thus, for a solution to the \textit{short-circuit} problem (III.A), 
it must be checked  that $\varphi (0) = \beta_1 b_1 + \beta_2 b_2$
and $t_{32} (0) =  \beta_1 e_{14} f_1 + \beta_2 e_{14} f_2$ are indeed 
equal to zero.
These conditions are equivalent to checking that 
\begin{equation}
\begin{vmatrix} 
  b_1 & b_2 \\
  f_1  & f_2  
\end{vmatrix} =0.
\end{equation}

For a solution to the \textit{open-circuit} problem (III.B), 
it must be checked 
that $t_{32} (0) =  \beta_1 e_{14} f_1 + \beta_2 e_{14} f_2$
and 
$d_2 (0)=\beta_1 (e_{14}^2/c_{44})g_1 + \beta_2 (e_{14}^2/c_{44})g_2$ 
are equal to zero.
These conditions are equivalent to checking that 
$ \begin{vmatrix} 
  f_1 & f_2 \\
  g_1  & g_2  
\end{vmatrix} =0$.
As expected\cite{BaLo76, BrGi79}, this condition is never satisfied.

For a solution to the \textit{free boundary} problem (III.C), 
it must be checked 
that $t_{32} (0) =  \beta_1 e_{14} f_1 + \beta_2 e_{14} f_2 =0$
and that $d_2 (0) = i\epsilon_0 \varphi(0)$ that is,
$\beta_1 (e_{14}^2/c_{44})g_1 + \beta_2 (e_{14}^2/c_{44})g_2 
= i\epsilon_0 (\beta_1 b_1 + \beta_2 b_2)$.
These conditions are equivalent to checking that 
\begin{equation} \label{checkFree}
\begin{vmatrix} 
  f_1 & f_2 \\
  \dfrac{e_{14}^2}{c_{44}\epsilon_0} g_1 - i b_1 
 &  \dfrac{e_{14}^2}{c_{44}\epsilon_0} g_2 - i b_2  
\end{vmatrix} =0.
\end{equation}
In general, this condition is met only for a limited range of the 
  cut angle $\theta$.

Finally, for a solution to the \textit{thin conducting layer boundary} 
problem (III.D), it must be checked 
that $t_{32} (0) =  0$
and that $d_2 (0) = (\epsilon + i\epsilon_0) \varphi(0)$.
These conditions are equivalent to checking that 
\begin{equation}
\begin{vmatrix} 
  f_1 & f_2 \\
  \dfrac{e_{14}^2}{c_{44}(\epsilon_0 - i\epsilon)} g_1 - i b_1 
 &  \dfrac{e_{14}^2}{c_{44}(\epsilon_0 - i\epsilon)} g_2 - i b_2  
\end{vmatrix} =0.
\end{equation}

\section{Examples}

In this Section, the secular equations derived in III and the tests 
presented in IV are employed 
to find numerically the wave speed  and its range of existence 
for three crystals, one with cubic $\bar{4}3$m symmetry, 
one with tetragonal $\bar{4}2$m symmetry, and one with cubic 23 
symmetry.
Data collected from the specialized literature are used for the values 
of the mass densities, of the stiffnesses, and of the piezoelectric 
and dielectric constants. 
In order to graph the depth profiles, a frequency of 100 MHz and a 
mechanical displacement of $10^{-13}$ m at $x_2=0$ are picked, to fix 
the ideas.
Note that at a 45$^\circ$ angle of cut, the profiles present 
pure (non-oscillating) exponential decay, because the propagation 
condition \eqref{inhomPropCond} is a biquadratic 
and the corresponding roots are purely imaginary; 
they are essentially similar to those displayed by Bright and 
Hunt\cite{BrHu89}. 
Here the profiles are computed at angles $\ne 45^\circ$.

\subsection{AlAs}

For Aluminum Arsenide ($\bar{4}3$m symmetry) the physical quantities 
of interest are\cite{KiHu90}: $\rho = 3760$
kg$\cdot$m$^{-3}$, $c_{44} = 58.9 \times 10^{9}$ N$\cdot$m$^{-2}$, 
$\hat{e}_{14} = -0.225$ C$\cdot$m$^{-2}$, and
$\epsilon_{11} = 10.06 \epsilon_0$.

Using the results of III and IV, it is found that the speeds of the 
piezoacoustic SH surface wave with metallized (III.A) and with free 
(III.C) boundary conditions are almost indistinguishable on a graph 
from the speed of the bulk shear wave. 
For instance at $45^\circ$, these speeds are (m$\cdot$s$^{-1}$): 
3976.784, 3976.965, and 3976.966, respectively.
Note however that the SH surface wave for the metallized (``shorted") 
boundary condition exists for 
all values of $\theta$ (within the range delimited above by the speed 
at 45$^\circ$ and below by the speed at $0^\circ$ and at $90^\circ$, 
which is 3957.890 m$\cdot$s$^{-1}$) whereas the SH surface wave for 
the un-metallized (``free'') boundary condition exists only within a 
limited range, delimited above by the speed at 45$^\circ$ 
and below by the speed at $45.0^\circ \pm 8.51^\circ$, 
which is 3975.276 m$\cdot$s$^{-1}$.

Figure 2 displays the variations of the three speeds as a function 
of $\theta$.
The speed of the bulk shear wave is always above the speed of the 
SH surface wave for the metallized boundary condition; 
they are both defined everywhere. 
The speed of the SH surface wave for the un-metallized 
boundary condition is intermediate between these two speeds, 
but exists only in the range [$36.49^\circ, 53.51^\circ$]. 
A zoom is provided  for this range.
In that zoom, the curve for the bulk shear wave almost coincides 
with the curve for the SH surface wave corresponding to the 
un-metallized boundary condition; together they form the upper curve 
whilst the lower curve represents the variations of the SH surface
wave speed corresponding to the metallized boundary condition.

\begin{figure}
\begin{centering}
\epsfig{figure=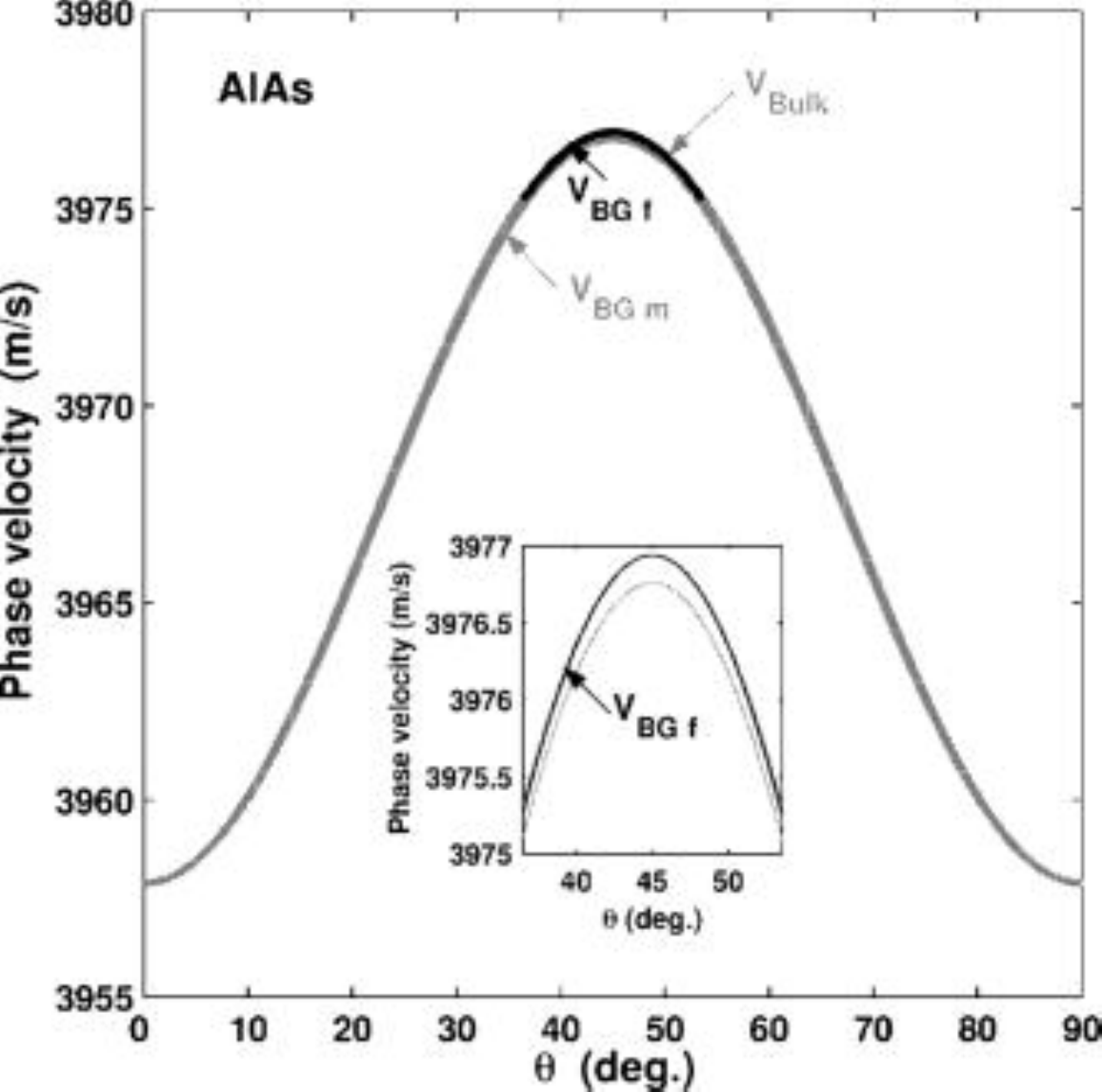, width=0.5\textwidth}
\caption{Speeds of piezoacoustic waves in a AlAs
($\bar{4}3$m symmetry) crystal, as a function of the cut angle: 
bulk shear wave (upper curve), SH surface wave for free 
(un-metallized) boundary conditions (intermediate curve), and 
SH surface wave for metallized boundary conditions (lower curve).}
\end{centering}
\end{figure}

Note that the simple test for the solution's validity presented in 
IV.B works perfectly here and henceforward. 
Thus using a 40 digit precision under MAPLE for AlAs, the modulus of 
the determinant in \eqref{checkFree} is found to be less than 
10$^{-19}$ at 45.0$^\circ \pm 8.5125^\circ$ and more than 0.2 at 
$45.0^\circ \pm 8.5129^\circ$.

Figure 3 shows the variations  with depth of the fields of 
interest (mechanical displacement, shear stress, electrical potential, 
 electric induction) for the  SH surface wave corresponding to the 
metallized boundary condition at $\theta = 22^\circ30'$.
The variations of the fields are presented over 250 wavelengths, and 
zooms are provided for the [0, 5] wavelengths range, where $\phi$, 
$\sigma_{32}$, and $D_2$ undergo rapid changes.

\begin{figure}
\begin{centering}
\epsfig{figure=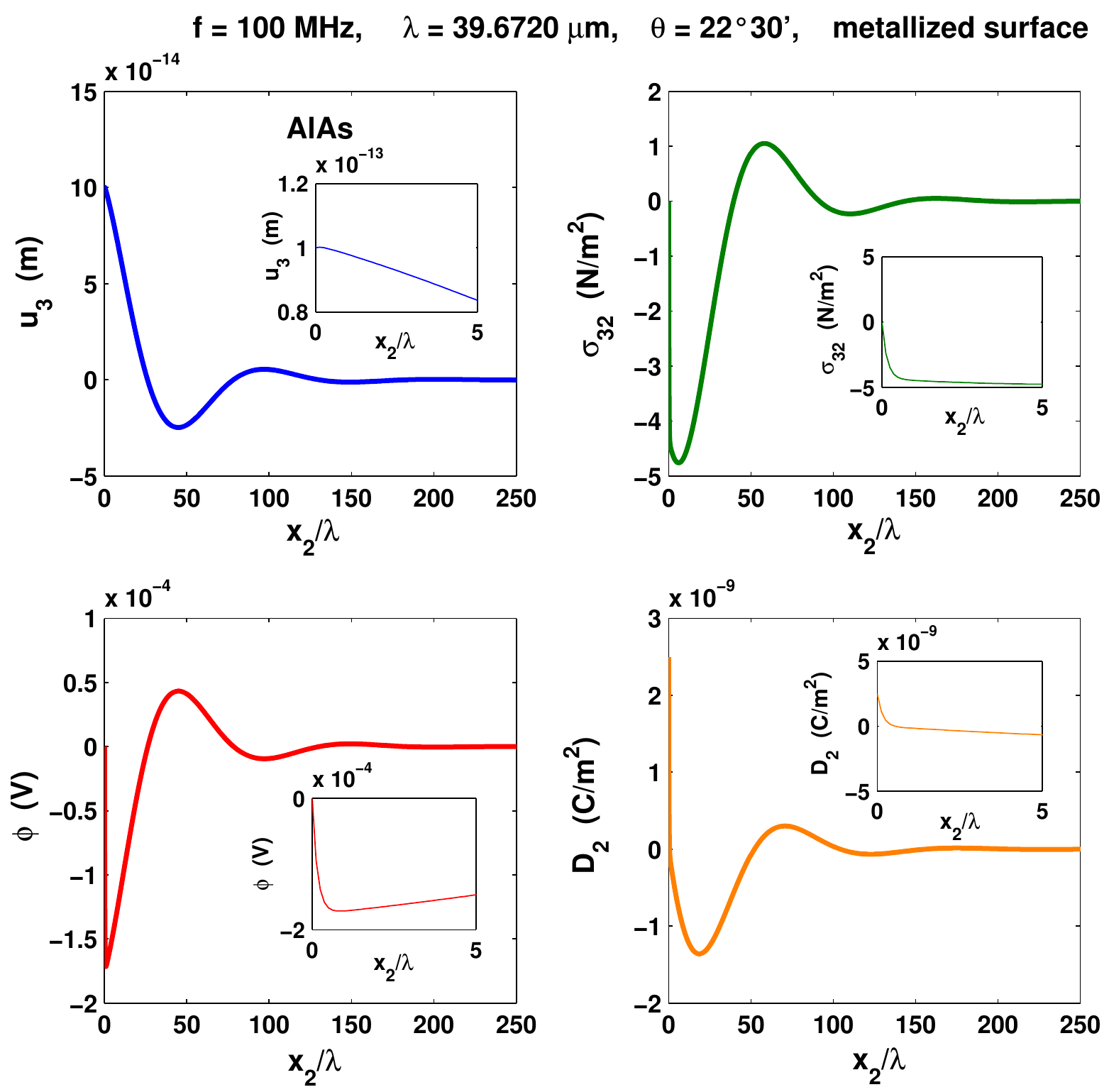, width=0.7\textwidth}
\caption{Depth profiles of the SH surface wave for metallized boundary 
conditions in a AlAs crystal cut at $22^\circ30'$: mechanical 
displacement, shear stress, electrical potential, electric induction.}
\end{centering}
\end{figure}

\subsection{BLGO}

Soluch \textit{et al.}\cite{SKPB85} measured experimentally the 
elastic, piezoelectric, and dielectric properties of BaLaGa$_3$O$_7$ 
 ($\bar{4}2$m symmetry, mass density: $\rho = 5450$ kg$\cdot$m$^{-3}$) 
as: 
$c_{44} = 39 \times 10^9$ N$\cdot$m$^{-2}$, 
$\hat{e}_{14} = 0.29$ C$\cdot$m$^{-2}$, and
$\epsilon_{11} = 12.4 \epsilon_0$.

Here, the speeds of the piezoacoustic SH surface wave
with metallized (III.A) and with free (III.C) boundary conditions 
differ more notably than in the previous example from the speed of the 
bulk shear wave. 
For instance at $45^\circ$, these speeds are (m$\cdot$s$^{-1}$): 
2700.739, 2701.239, and 2701.242, respectively.
The range of values for the wave speed $v_{\text{BGm}}$ of the 
metallized  ``shorted" boundary condition is delimited above by the 
speed at 45$^\circ$ and below by the speed at $0^\circ$ and at 
$90^\circ$, which is 2675.063 m$\cdot$s$^{-1}$. 
For the un-metallized ``free'' boundary condition, 
the corresponding (limited) range 
for the wave speed $v_{\text{BGf}}$ is bounded above by the speed at 
45$^\circ$ and below by the speed at $45.0^\circ \pm 7.485^\circ$, 
which is 2699.369 m$\cdot$s$^{-1}$.
The difference between the two speeds is the largest at 
$\theta = 45^\circ$; there the ratio 
$2(v_{\text{BGf}}- v_{\text{BGm}})/v_{\text{BGf}}$ is equal to 3.70 
$\times 10^{-4}$. 

Figure 4 displays the variations  with $\theta$ of the speeds for the 
bulk shear wave, for the SH surface wave corresponding to the 
un-metallized boundary condition, and for the SH surface wave 
corresponding to the metallized boundary condition.
Figure 5 shows the variations  with depth of the fields of 
interest (mechanical displacement, shear stress, electrical potential, 
electric induction) for the SH surface wave corresponding to the 
metallized boundary condition at $\theta = 22^\circ30'$.
Similar comments to those made for Figure 3 apply.

\begin{figure}
\begin{centering}
\epsfig{figure=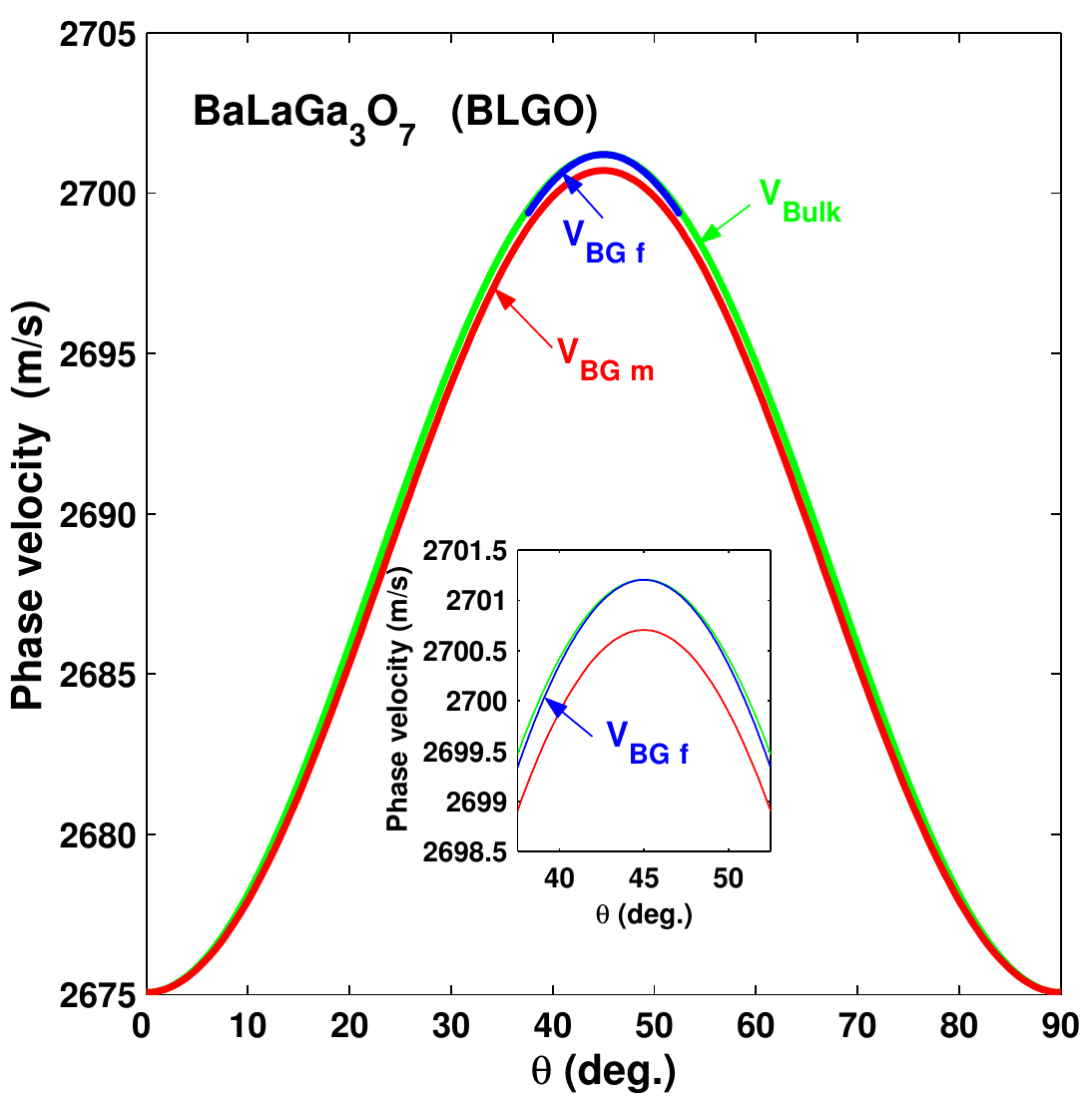, width=0.5\textwidth}
\caption{Speeds of piezoacoustic waves in a BaLaGa$_3$O$_7$ 
($\bar{4}2$m symmetry) crystal, as a function of the cut angle: 
bulk shear wave (upper curve), SH surface wave for free 
(un-metallized) boundary conditions (intermediate curve), and 
SH surface wave for metallized boundary conditions (lower curve).
}
\end{centering}
\end{figure}

\begin{figure}
\begin{centering}
\epsfig{figure=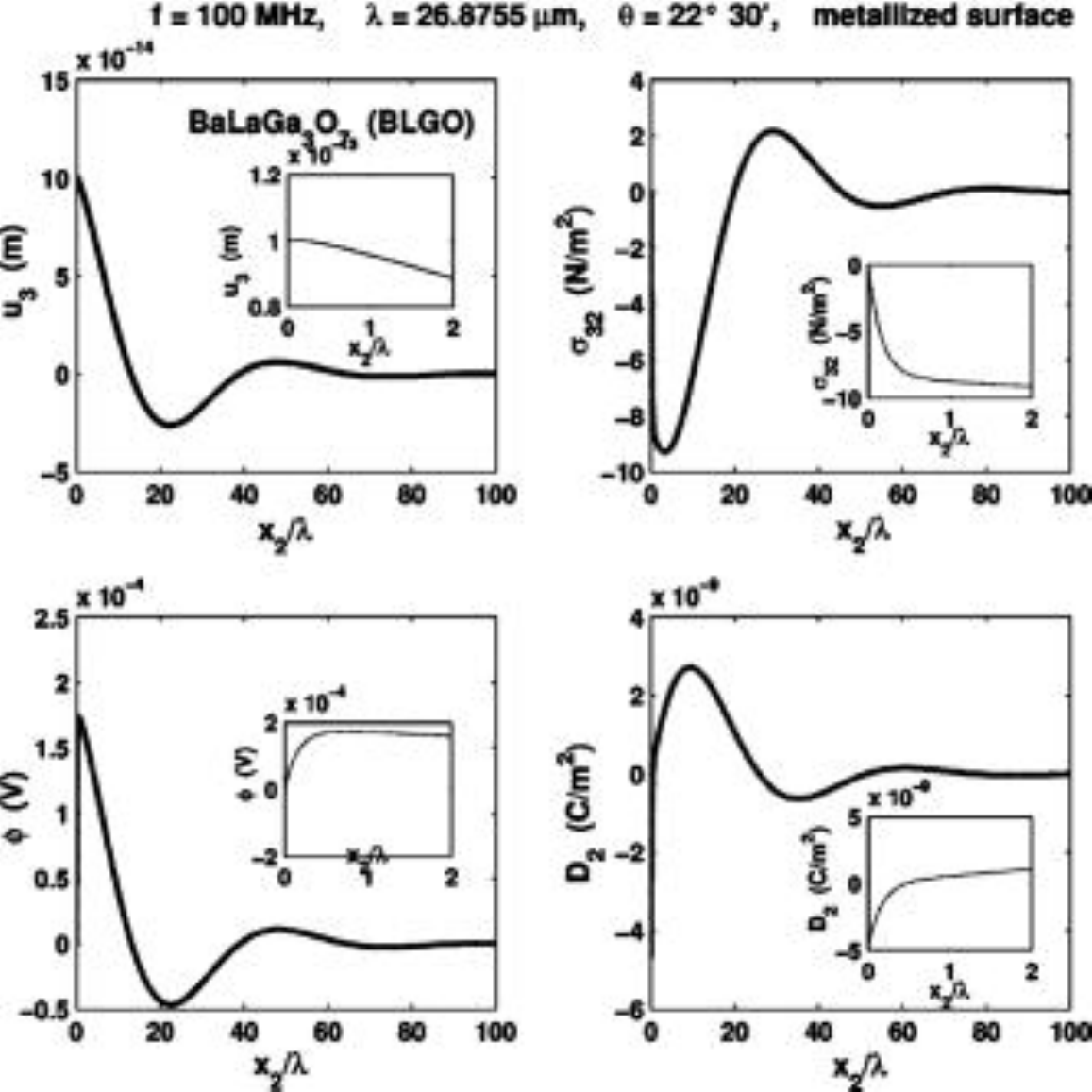, width=0.5\textwidth}
\caption{Depth profiles of the piezoacoustic SH surface wave 
for metallized boundary conditions in a BaLaGa$_3$O$_7$ crystal cut at 
$22^\circ30'$: mechanical displacement, shear stress, electrical 
potential, electric induction}
\end{centering}
\end{figure}

\subsection{BGO}

The relevant physical quantities of Bismuth Germanium Oxide 
(Bi$_{12}$GeO$_{20}$, 23 symmetry) are\cite{Auld73}: $\rho = 9200$
kg$\cdot$m$^{-3}$, $c_{44} = 25.52 \times 10^{9}$ N$\cdot$m$^{-2}$, 
$\hat{e}_{14} = 0.983$ C$\cdot$m$^{-2}$, and
$\epsilon_{11} = 38.0 \epsilon_0$.

Here the differences between the speeds of the bulk shear wave, 
of the SH surface wave corresponding to the metallized boundary 
condition, and of the SH surface wave corresponding to the 
un-metallized boundary condition are more marked than in the previous 
example. 
At $45^\circ$, these speeds are (m$\cdot$s$^{-1}$): 
1747.812,  1756.836, and 1756.846, respectively. 
At $0^\circ$ and at $90^\circ$, the speeds of the bulk shear wave and 
of the SH surface wave corresponding to the metallized boundary 
condition are both equal to 1665.507 m$\cdot$s$^{-1}$.
The SH surface wave for the un-metallized boundary condition exists 
only in the range $45.00 \pm 3.149^\circ$, 
and at the extremities of this range, its speed is 1755.068 
m$\cdot$s$^{-1}$.

Figure 6 displays the variations of the three wave speeds as a 
function of $\theta$.
Figure 7  shows the variations of the quantity 
$2(v_{\text{BGf}}- v_{\text{BGm}})/v_{\text{BGf}}$ 
with the angle of cut; its largest (smallest) value is 
1.027$\times10^{-2}$ at $45.00^\circ$ (0.975$\times10^{-2}$ at 
$48.149^\circ$).

\begin{figure}
\begin{centering}
\epsfig{figure=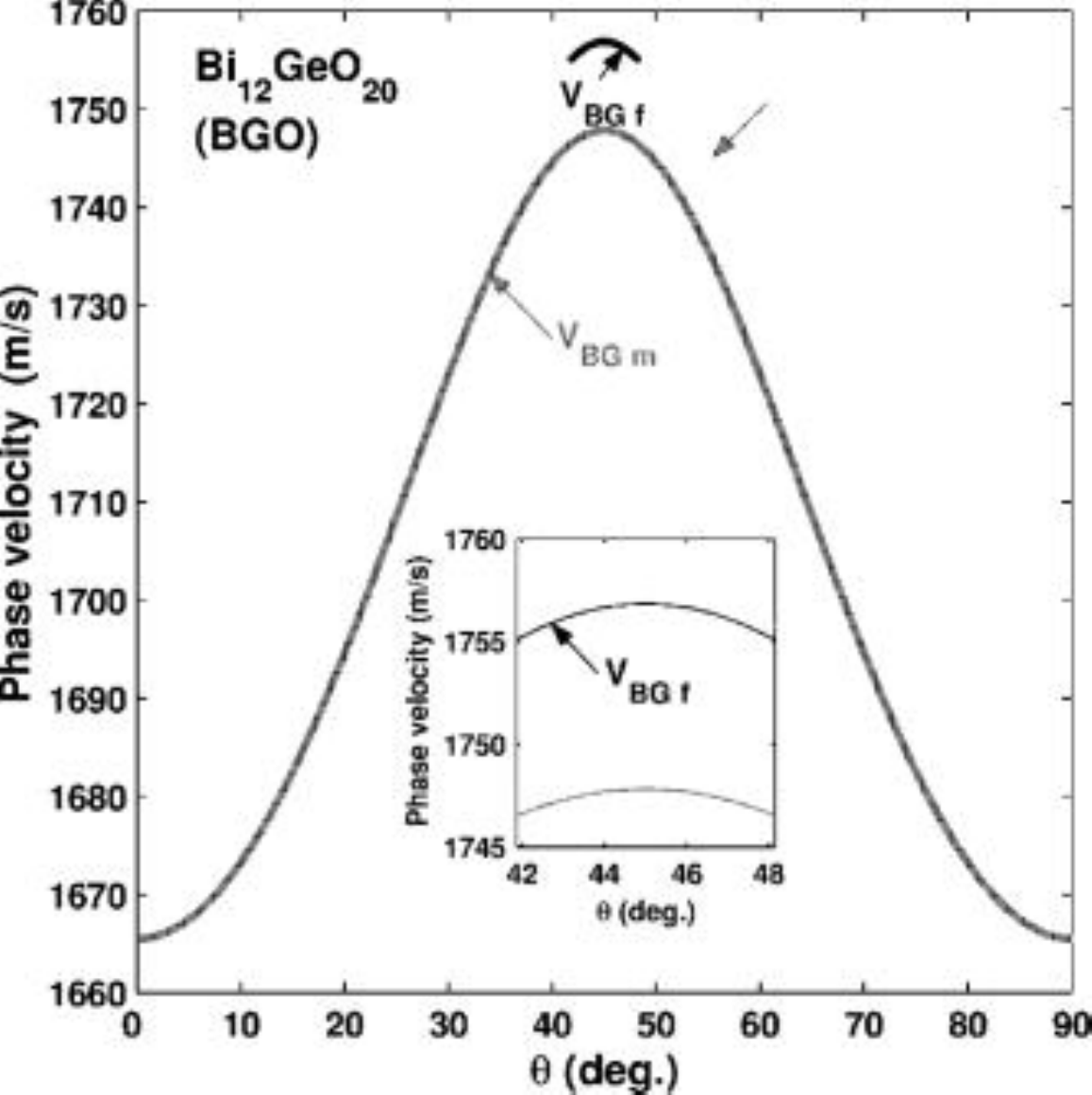, width=0.5\textwidth}
\caption{Speeds of piezoacoustic waves in a Bi$_{12}$GeO$_{20}$
(23 symmetry) crystal, as a function of the cut angle: 
bulk shear wave (upper curve), SH surface wave for free 
(un-metallized) boundary conditions (intermediate curve), and 
SH surface wave for metallized boundary conditions (lower curve).}
\end{centering}
\end{figure}

\begin{figure}
\begin{centering}
\epsfig{figure=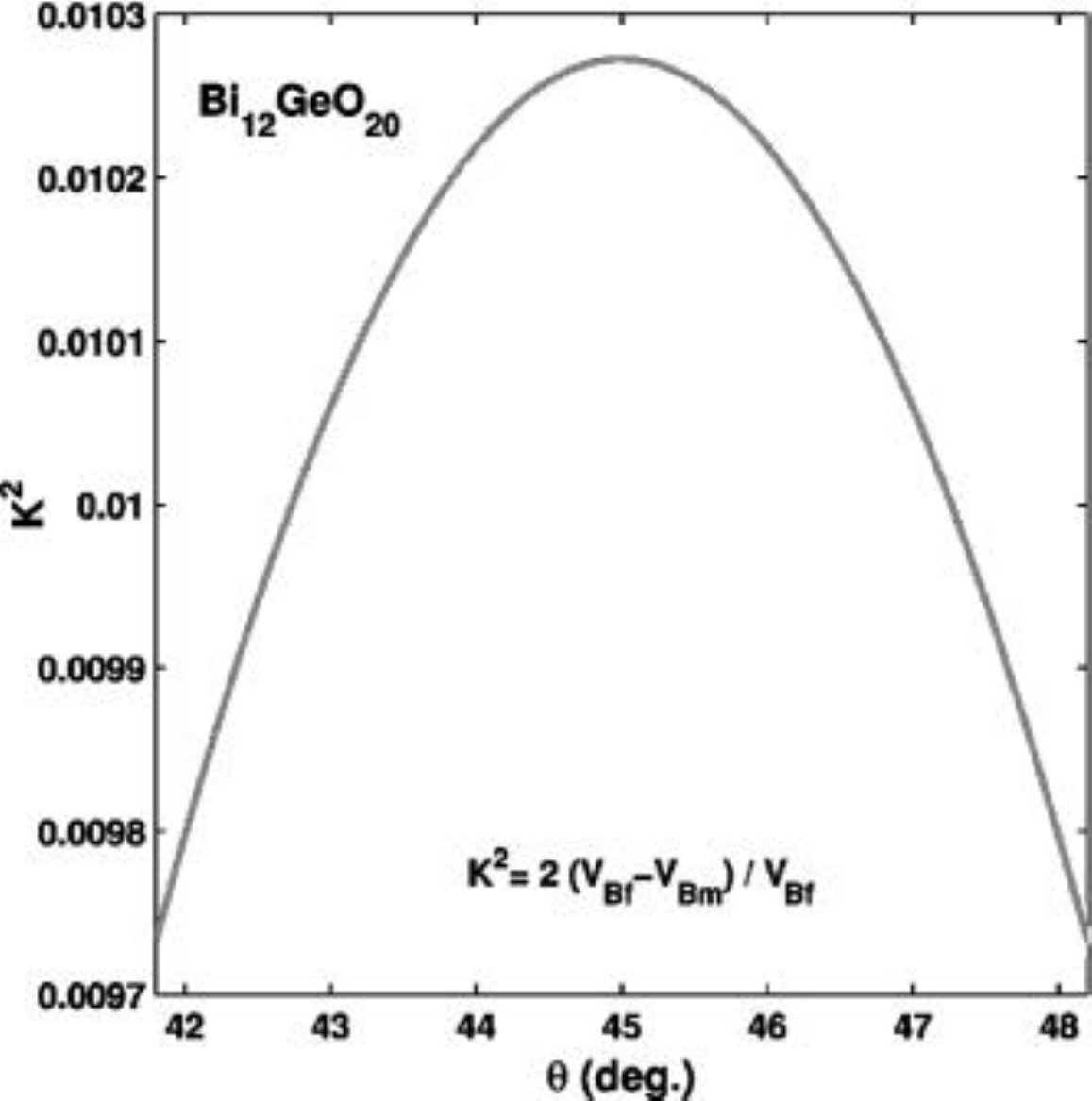, width=0.5\textwidth}
\caption{Variations of 
    $2(v_{\text{BGf}}- v_{\text{BGm}})/v_{\text{BGf}}$ with $\theta$ 
in a Bi$_{12}$GeO$_{20}$ (23 symmetry) crystal.}
\end{centering}
\end{figure}

Figure 8 shows the variations with depth of the fields of 
interest (mechanical displacement, shear stress, electrical potential, 
 electric induction) for the SH surface wave corresponding to the 
metallized boundary condition at $\theta = 22^\circ30'$.
Similar comments to those made for Figures 3 and 4 apply.

\begin{figure}
\begin{centering}
\epsfig{figure=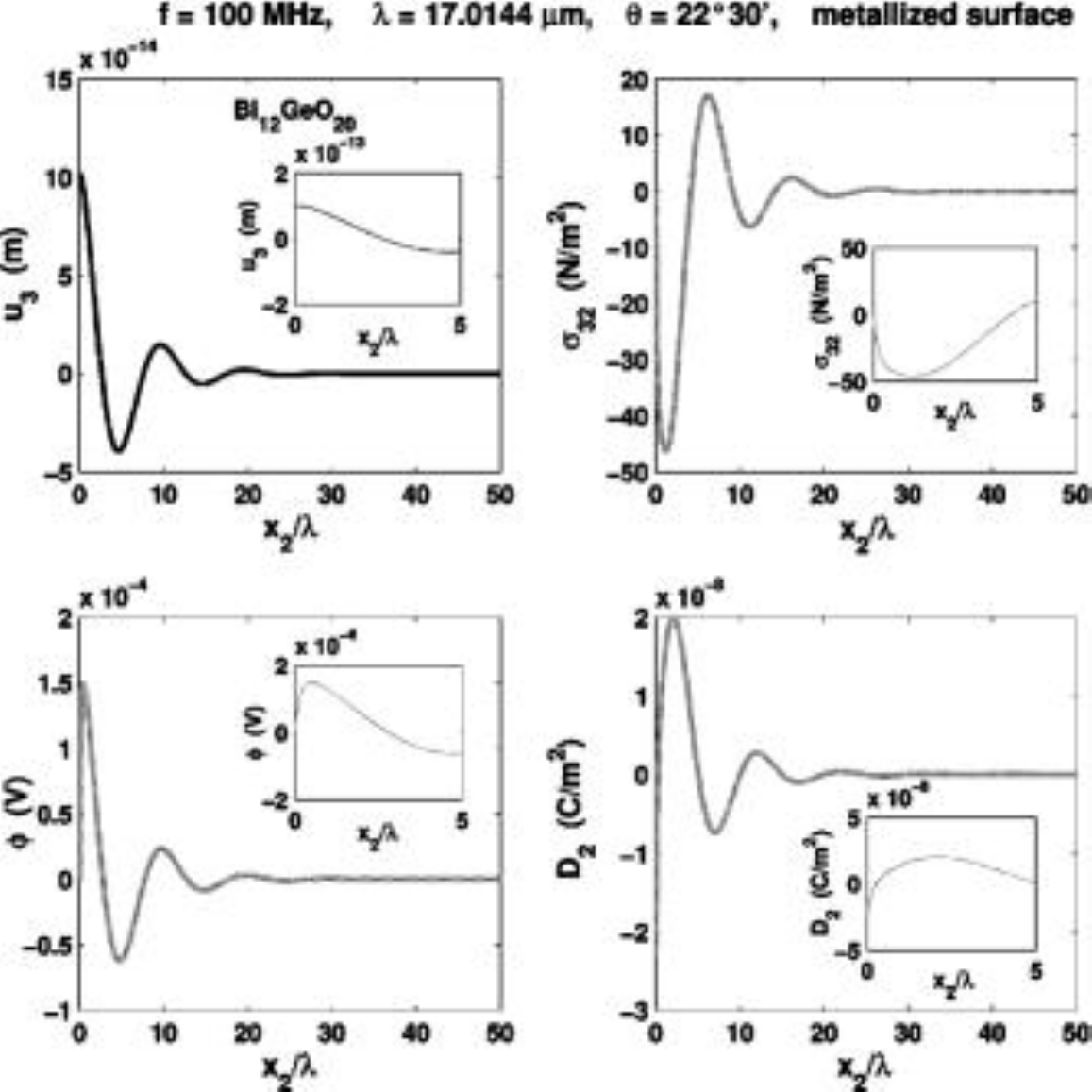, width=0.5\textwidth}
\caption{Depth profiles of the piezoacoustic SH surface wave for metallized 
boundary conditions in a Bi$_{12}$GeO$_{20}$ crystal cut at 
$22^\circ30'$: 
mechanical displacement, shear stress, electrical potential, 
electric induction.}
\end{centering}
\end{figure}

Figure 9 shows the variations  with depth of the same fields 
for the SH surface wave corresponding to the un-metallized 
boundary condition at $\theta = 42^\circ30'$.
By comparison with the previous Figure it can be seen 
that the SH surface wave for the un-metallized boundary condition 
penetrates far more deeply than the SH surface wave for the metallized 
boundary condition.
The electrical potential and the electric induction are plotted inside 
the crystal for the range [0, 500] wavelengths, and also in the 
vacuum over the crystal for the range [-10, 0] wavelengths; 
the continuity of these fields across the interface is made apparent 
with a zoom for the range [-5, 5] wavelengths.

\begin{figure}
\begin{centering}
\epsfig{figure=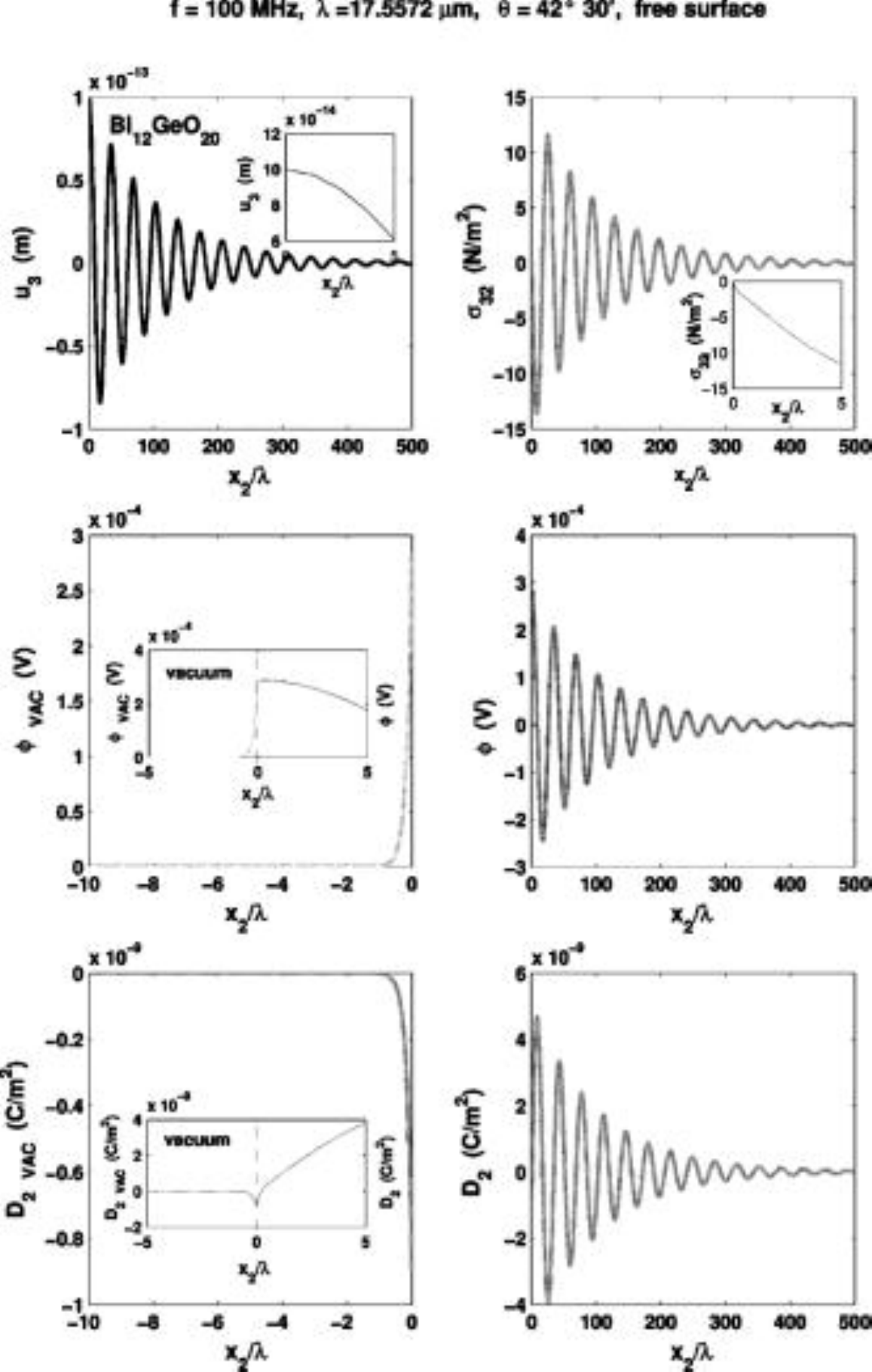, width=0.8\textwidth}
\caption{Depth profiles of the piezoacoustic SH surface wave for free 
(un-metallized) boundary conditions in a Bi$_{12}$GeO$_{20}$ crystal 
cut at  $42^\circ30'$: 
mechanical displacement, shear stress, electrical potential, 
electric induction.
}
\end{centering}
\end{figure}

\section{Concluding remarks}

The method of resolution for the title problem of this paper 
is based on the fundamental equations \eqref{equations}.
This method has proved itself to be very effective and versatile.
The end result is the complete analytical elucidation of the problem, 
for a great variety of surface impedance problems in a piezoelectric 
half-space i.e., problems where the electric induction is 
proportional to the electrical potential: $k\Phi = iv ZD_2$ (say) at 
the boundary plane. 
The method can be followed through when the impedance $Z$ is zero 
(short-circuit), infinite (open-circuit), pure imaginary 
(free boundary), or complex (thin  conducting layer). 
It has already been used for other types of surface impedance problems 
for elastic interface waves (Stoneley waves\cite{Dest03d,Dest04b}, 
Scholte waves\cite{Dest04a}) and could be adapted to 
configurations\cite{Kali67} with a resistance force proportional 
to the normal velocity, a mass concentrated in a thin surface layer,
a system of elastic oscillators resting on an elastic half-space, 
a thin elastic layer longitudinally deformable, etc.
The method can also accommodate a coupling between elastic and 
piezoelectric fields, in situations such as the one treated here or 
for instance, the case of interface acoustic waves at a  domain 
boundary\cite{MoWe98}.

Here, attention was restricted to Bleustein-Gulyaev waves in 
tetragonal $\bar{4}$ piezoelectric crystals. 
The extension to the classes of orthorhombic 222 or monoclinic 2 
crystals is straightforward and only requires the computation of the 
elements of the matrix $\mathbf{N}$ in Eq.~\eqref{motion}. 
Their general expression is given for instance by Abbudi and 
Barnett\cite{AbBa90}. 
As an illustration, they are now presented for rhombic 222 crystals.

For such a crystal, the relevant non-zero piezoacoustic constants in 
the crystallographic coordinate system are 
$\hat{c}_{44}$, $\hat{c}_{55}$, $\hat{e}_{14}$, $\hat{e}_{25}$, 
$\hat{\epsilon}_{11}$, and $\hat{\epsilon}_{22}$.
In the coordinate system obtained  after the rotation 
Eq.~\eqref{rotation}, they are 
\begin{align} 
&     c_{44} = \hat{c}_{44}\cos^2 \theta + \hat{c}_{55}\sin^2 \theta, 
  &&   c_{45} = (\hat{c}_{44} - \hat{c}_{55})\cos \theta \sin \theta,
\nonumber \\ 
&    c_{55} = \hat{c}_{55}\cos^2 \theta + \hat{c}_{44}\sin^2 \theta, 
&&  \epsilon_{12} 
= (\hat{\epsilon}_{22} - \hat{\epsilon}_{11})\cos \theta \sin \theta, 
\nonumber \\ 
&     \epsilon_{11} 
     = \hat{\epsilon}_{11} \cos^2 \theta 
              + \hat{\epsilon}_{22} \sin^2 \theta, 
&&   \epsilon_{22} = \hat{\epsilon}_{22} \cos^2 \theta 
                          + \hat{\epsilon}_{11} \sin^2 \theta, 
\nonumber \\ 
&   e_{14} = \hat{e}_{14} \cos^2 \theta  - \hat{e}_{25} \sin^2 \theta, 
&& e_{15} = (\hat{e}_{14} + \hat{e}_{25})\cos \theta \sin \theta,
\nonumber \\ 
&   e_{25} = \hat{e}_{25} \cos^2 \theta  - \hat{e}_{14} \sin^2 \theta,
&& e_{24} = - \hat{e}_{15}.
\end{align}
The equations of motion can be cast in the form Eq.~\eqref{motion}
where the matrix $\mathbf{N}$ is defined by its $2 \times 2$ 
blocks $\mathbf{N^{(1)}_1}$, $\mathbf{N^{(1)}_2}$, 
$\mathbf{K^{(1)}}$ in Eq.~\eqref{structure}.
The components of $-\mathbf{N^{(1)}_1}$ and 
$\mathbf{N^{(1)}_2}$ are given here by 
\begin{equation}
 \begin{bmatrix} 
 \left(  \dfrac{\epsilon_{22}c_{45}}{e_{14}e_{15}}
            - \dfrac{e_{25}}{e_{14}}     \right)\kappa^2 
  &\left(  \dfrac{\epsilon_{22}}{e_{15}}
             + \dfrac{\epsilon_{12}}{e_{14}}     \right)\kappa^2 
\vspace{8pt} \\
 -\left(  \dfrac{c_{44}e_{25}}{e_{14}e_{15}}
            + \dfrac{c_{45}}{e_{14}}     \right)\kappa^2 
  &\left(  \dfrac{\epsilon_{12}c_{44}}{e_{14}e_{15}}
             -1     \right)\kappa^2 
 \end{bmatrix},
\quad 
\begin{bmatrix} 
   \dfrac{\epsilon_{22}}{e_{14}e_{15}} \kappa^2 
  & - \dfrac{1}{e_{14}} \kappa^2 
\vspace{8pt} \\
 - \dfrac{1}{e_{14}} \kappa^2 
  & - \dfrac{c_{44}}{e_{14}e_{15}}\kappa^2 
 \end{bmatrix},
\end{equation} 
respectively, and those of $\mathbf{K^{(1)}}$ are
\begin{align}
& K^{(1)}_{11} =  X - c_{55}
 - \left( c_{44} \dfrac{e_{25}^2}{e_{14}e_{15}}
            + 2c_{45} \dfrac{e_{25}}{e_{14}}
              - c_{45} \dfrac{\epsilon_{22}c_{45}}{e_{14}e_{15}} 
       \right)\kappa^2, 
\nonumber  \\
& K^{(1)}_{12} =  - e_{15}
     + \left[  \dfrac{c_{44}\epsilon_{12}}{e_{14}}
             + \dfrac{c_{45}\epsilon_{22}}{e_{15}}
    +  e_{25} \left(\dfrac{\epsilon_{12}c_{44}}{e_{14}e_{15}}-1\right)
        \right] \kappa^2, 
\nonumber  \\
& K^{(1)}_{22} =  - \epsilon_{11}
     + \left[   
  \epsilon_{12} 
  \left(\dfrac{\epsilon_{12}c_{44}}{e_{14}e_{15}}-2\right)
   - \epsilon_{22} \dfrac{e_{14}}{e_{15}}
          \right] \kappa^2. 
\end{align} 
Here, the quantity $\kappa^2$ is defined by
\begin{equation} 
\kappa^2 = \dfrac{e_{14}e_{15}}{\epsilon_{11}c_{44} + e_{15}^2}. 
\end{equation} 

Finally, in guise of a Conclusion, the main relevant advances 
toward the full resolution of the problem presented in the paper are
recapitulated.
In the purely elastic case, the secular equation for Rayleigh waves 
polarized in a plane of symmetry was derived by Currie\cite{Curr79} 
who, using an algebraic approach based on the Stroh formalism, 
obtained the equations,
\begin{equation} 
\overline{\mathbf{U}}(0) \cdot \mathbf{K^{(n)}} \mathbf{U}(0) = 0, 
\end{equation} 
where $\mathbf{U}(0)$ is the mechanical displacement on the free 
surface. 
Although these equations are also valid in generally anisotropic 
crystals, his derivation of the secular equation for triclinic 
(no symmetry) crystals apparently leads to a trivial identity. 
This problem was later corrected by Taylor and Currie\cite{TaCu81} 
and by Taziev\cite{Tazi89} (see also Ting\cite{Ting04a}). 
In contrast to these approaches based on the formulation of the 
equations of motion as a first-order differential system for the 
displacement-traction vector, Mozhaev\cite{Mozh95} wrote the 
equations of motion as a second-order differential system for the 
displacement vector, 
\begin{equation} \label{2ndOrder1}
\mathbf{\alpha U}'' - i\mathbf{\beta U}' + \mathbf{\gamma U}
 = \mathbf{0}, 
\end{equation}
where $\mathbf{\alpha}$, $\mathbf{\beta}$, $\mathbf{\gamma}$, 
are real symmetric matrices. 
Then, using first integrals, he quickly derived the secular equation 
for orthorhombic crystals. 
Destrade\cite{Dest01} rewrote the equations of motion, this time in 
the form 
\begin{equation} \label{2ndOrder2}
\mathbf{\hat{\alpha} t}'' - i\mathbf{\hat{\beta} t}' 
  + \mathbf{\hat{\gamma} t} = \mathbf{0}, 
\end{equation}
for the tractions, where $\mathbf{\hat{\alpha}}$, 
$\mathbf{\hat{\beta}}$, $\mathbf{\hat{\gamma}}$, 
are real symmetric matrices. 
Adapting Mozhaev's first integrals, he re-derived (unaware of Currie's 
result) the secular equation for Rayleigh waves polarized in a 
symmetry plane. 
He also mentioned (and the proof was later given in the 
review article by Ting\cite{Ting04b}) that the method of first 
integrals could not be used for arbitrary anisotropy when the 
equations of motion are written as Eq.~\eqref{2ndOrder1} or 
Eq.~\eqref{2ndOrder2}.
Recently\cite{Dest03d, Dest04a, Dest04b} he made the connection 
between Currie's and Taziev's use of integer powers of the 
Stroh matrix $\mathbf{N}$ and Mozhaev's first integrals, as shown also 
here in II.C. 
Note that Mozhaev and Weihnacht\cite{MoWe02} were able to solve 
the problem of  SH surface modes of a 2mm crystal 
using first integrals of the piezoacoustic equations written as a 
second-order differential system.




\end{document}